\documentclass[twocoloumn]{article}
\usepackage{graphicx}
\usepackage{gc}
\heads{E.E. Kholupenko, A.V. Ivanchik and D.A. Varshalovich}
      {Distortion of the background radiation spectrum}

\def\la{\;\raise0.3ex\hbox{$<$\kern-0.75em\raise-1.1ex\hbox{$\sim$}}\;}
\def\ga{\;\raise0.3ex\hbox{$>$\kern-0.75em\raise-1.1ex\hbox{$\sim$}}\;}

\begin{document}
\twocolumn[
\Arthead{11}{2005}{1-2}{161}{165}

\Title{CMBR distortion concerned with recombination 
       \yy of the primordial hydrogen plasma}

\Author{E.E. Kholupenko\foom 1, 
        A.V. Ivanchik\foom 2, 
    and D.A. Varshalovich\foom 3} 
    {Ioffe Physical Technical Institute St.-Petersburg} 

\Abstract
    {CMBR distortion concerned with
recombination of the primordial plasma is calculated in frequency
band from 1 GHz to 100 GHz in the frame of the standard
cosmological model for different values of cosmological density
parameters: nonrelativistic matter density $\Omega_m$ and baryonic
matter density $\Omega_b$. Comparison of these results with
observational data which will be obtained from planned experiments
may be used for independent determination of the cosmological
parameters $\Omega_m$ and $\Omega_b$. }] 

\email 1 {eugene@astro.ioffe.ru}
\email 2 {iav@astro.ioffe.ru}
\email 3 {varsh@astro.ioffe.ru}

\section{Introduction}
The Universe was existing at the thermodynamical equilibrium
during 100 thousand
years after the electron~-~positron annihilation. When the temperature
of the Universe became low enough ($<2\cdot10^4$ K, it is corresponding 
to redshift $z<7000$) because of the cosmological
expansion, the recombination of the main elements began (Fig. 1).


\Figure{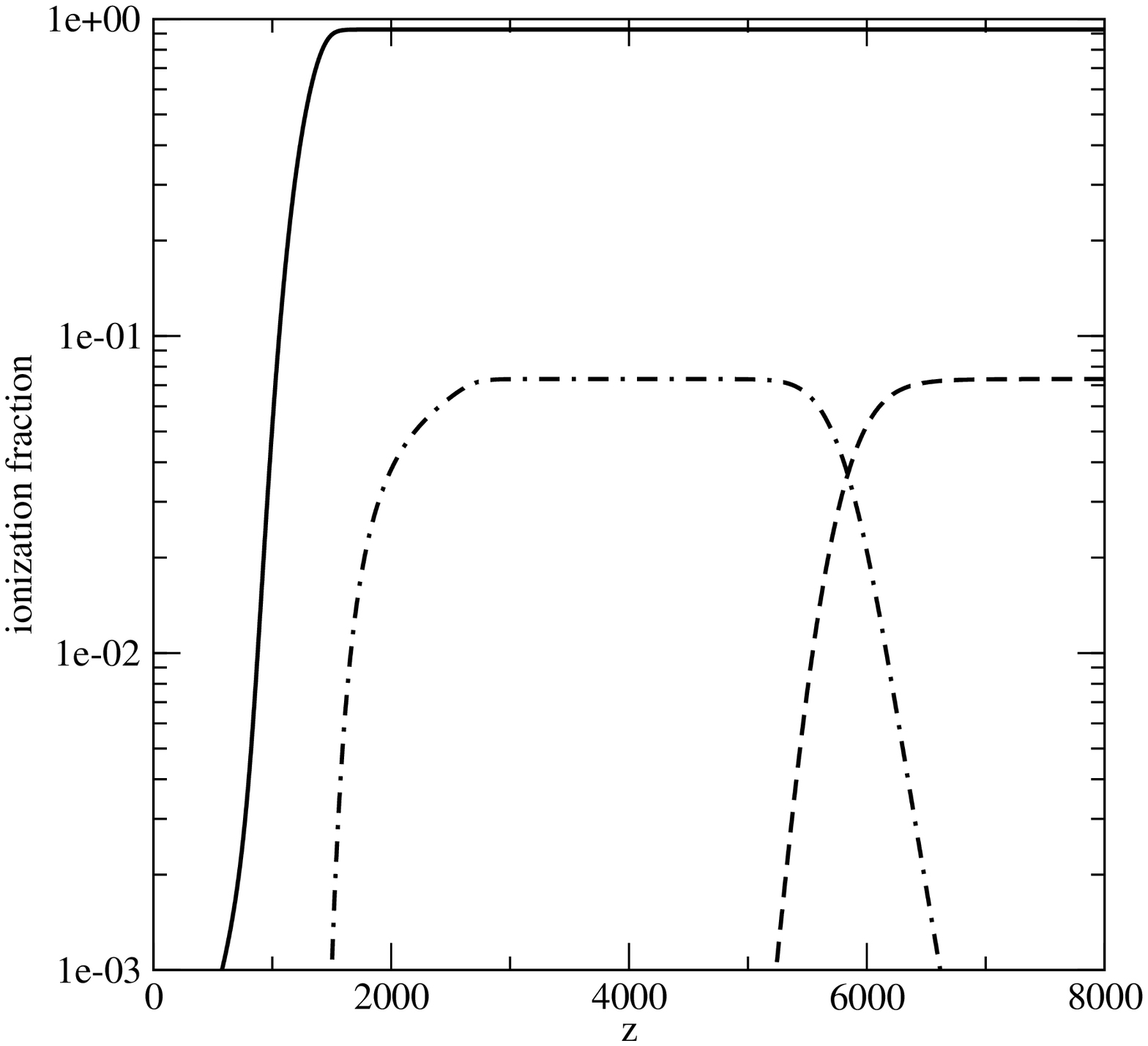} {The recombination of the main elements: 
the solid line corresponds to the recombination of HII, 
dashed -- HeIII, dashed-dot -- HeII}{540}{532}

This process leds to the birth of non-equilibrium photons. The
most of them had survived until the modern epoch and have appeared as
lines of the CMBR spectrum (Fig. 2).

\Figure{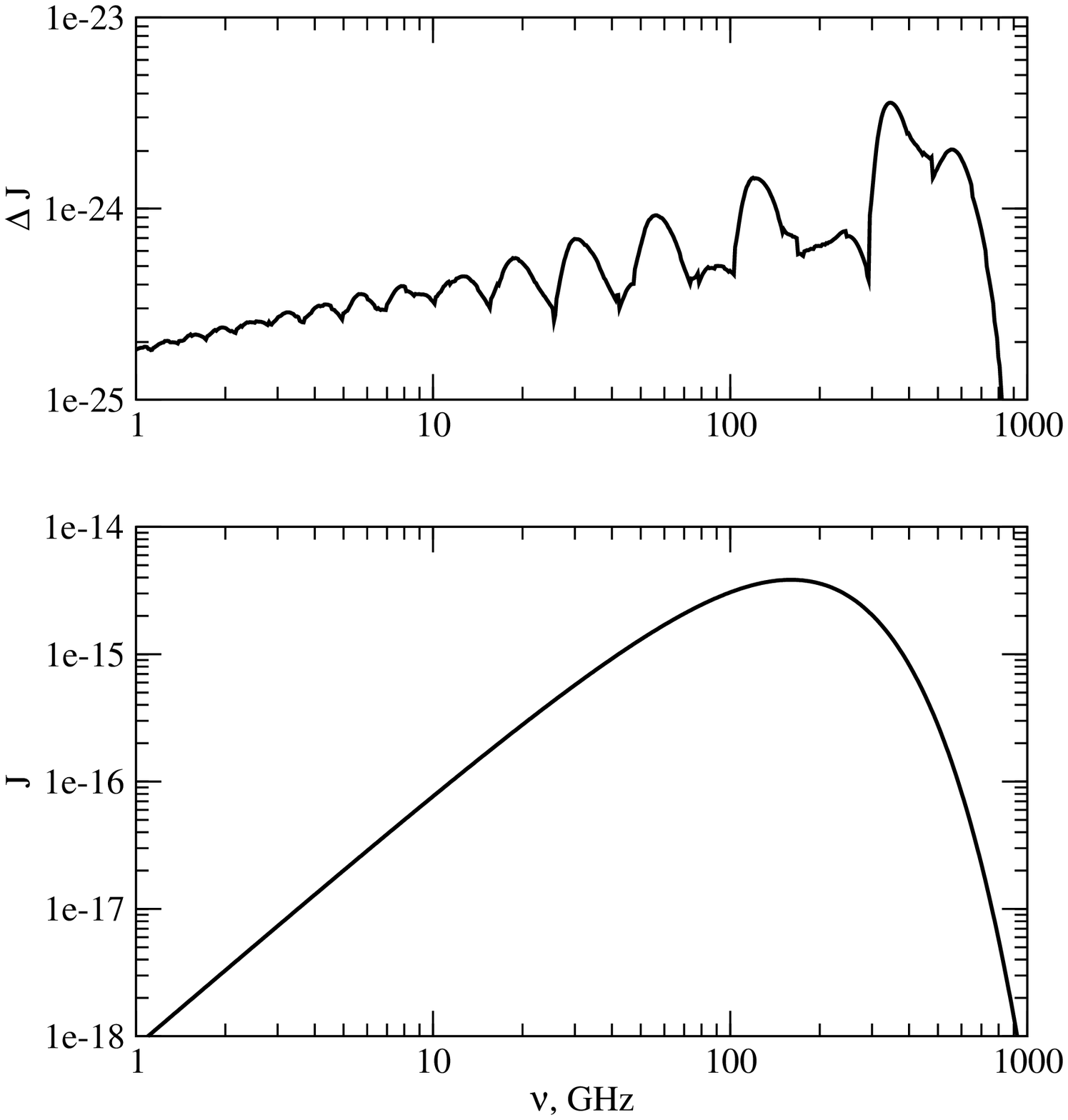} {Top picture: $\Delta J$ is the brightness 
[erg/cm$^{2}$/sec/Hz/ster]
of the non-equilibrium radiation concerned with the primordial hydrogen 
recombination. Bottom picture: $J$ 
[erg/cm$^{2}$/sec/Hz/ster] is the CMBR brightness as a 
function of frequency.}{540}{640}

The measurement of the intensity and the form of these lines would give
the information on the values of key cosmological parameters 
(e.g. $\Omega_b, \Omega_m$).

The calculation of the distortion of the CMBR spectrum requires:
\\1. The calculation of the level populations of the primordial
hydrogen atoms in the presence of a strong radiation
(photon per mode $\eta \gg 1$ for the transitions investigated)
which has the Planck spectrum at temperature decreasing with the cosmological time.
\\2. The modelling of the radiation transfer in the homogeneous expanding
environment.

\section{Physical model}
We used the standard cosmological model (see Tab.1).
The temperature, the concentration of protons, and the Hubble
constant as functions of redshift are:
\begin{equation}
T = T _0 (1+z), \;\;\;\; N_{tot} = N _{p0} (1+z) ^3,
\end{equation}
\begin{equation}
H = H_0 \sqrt{\Omega_\Lambda+\Omega_m (1+z)^3 + \Omega_{rel}(1+z)^4}
\label{gr0}
\end{equation}
\\ \bigskip
Table 1. The standard cosmological model parameters\\
\begin{tabular}{lll}
  \hline
  Value description & Symbol & Value \\
  \hline
  total matter& $\Omega_{tot}$ & ~~~~1 \\
  space curvature& $\Omega_k$ & ~~~~0 \\
  non-relativistic matter& $\Omega_m$ & $\sim 0.3$ \\
  baryonic matter& $\Omega_b$ & $\sim 0.04$ \\
  relativistic matter& $\Omega_{rel}$ & $\sim 10^{-4}$ \\
  vacuum-like energy& $\Omega_\Lambda$ & $\sim 0.7$\\
  Hubble constant & $H_0$ & 70 km/s/Mpc \\
  radiation temperature & $T_{0}$ & $2.725 \pm 0.002 $ K \\
  proton concentration & $N_{p0}$ & $\sim 2 \cdot 10^{-7}$ cm$^{-3}$ \\
  \hline
\end{tabular}
\bigskip \\
The characteristic values of the cosmological parameters in the
recombination epoch are presented in~Tab.~2. 
The knowledge of these values allows us to estimate the relative
CMBR temperature distortion concerned with the non-equilibrium radiation at
$i \rightarrow k$ transition ($h\nu_{ik}\ll k_BT$) at frequency
$\nu\simeq \nu_{ik}/(1+z)$ by the formula:

\begin{equation}
{\Delta T \over T}={c^3 \over 8 \pi \nu_{ik}^3}
{1 + z \over \Delta z}{h\nu_{ik} \over k_B T(z)}N_{tot}(z) \theta_{ik}
\label{estimate2}
\end{equation}
where $\Delta z$ is the duration of recombination epoch,
$\theta_{ik}$ is an element of the matrix, so-called ``matrix of
efficiency of radiative transitions'' (ERT-matrix was introduced 
by Bernstein et al. [1]). 
Their values are $10^{-2} - 10^{-4}$ for $i,k=10-40$
correspondingly (see [2]).
\\ \\ \bigskip
\begin{tabular}{lll}
Table 2\\
\hline
  Redshift & $z$ & 800 - 1600 \\
  Temperature & $T$ & 2200 - 4100 K \\
  Proton concentration & $N_p$ & 100 - 700 cm$^{-3}$ \\
\hline
\end{tabular}
\bigskip \\
For example, according to formula (\ref{estimate2}), the
relative distortion of the CMBR temperature concerned
with $17 \rightarrow 16$ transition (redshifted frequency
$\simeq 1.14$ GHz, $\theta\simeq 5\cdot 10^{-3}$) is
$\Delta T /T=2.5\cdot10^{-8}$.

 Formula (\ref{estimate2})
does not take into account the following facts:
\\1. There are overlaps of the line profiles because the cosmological
expansion leads to the formation of the wide line profile with
the relative frequency width about 0.3.
\\2. The variation of the recombination rate leads to the formation
of a non-rectangular line profile.
\\3. The elements of the ERT-matrix $\theta_{ik}$ depend on redshift.
They vary during recombination epoch within 20\%.

The points mentioned above made us to calculate
the formation of the CMBR distortion numerically.

\section{Initial equations}
\subsection{Description of population behavior}
We considered the hydrogen atoms in the excited states with
principal quantum number $n \ge 2$ ([1], [2]).
The transitions into the ground state are considered separately because
the optical depth for these transitions is much more than one ([3]).

For the description of the behavior of the hydrogen atoms we used
the quasi-stationary approximation of the kinetic equations averaged
over the angular momentum:
\begin{equation}
\sum_{k=1}^{\infty}A_{ki}x_k - \sum_{k=1}^{\infty}A_{ik}x_i
+\alpha_i N_{e} x_p - \beta_i x_i - J_i=0
\label{init_kin_eq1}
\end{equation}
where $x_i=N_i/N_{tot}$ is the relative population of the state with
principal quantum number $n=i+1$ (subscript $i=1$ corresponds to
the first excited state), $N_i$ is the population of state $i$
(the number of atoms in state $i$ per volume unit), $N_e$ is the free
electron concentration, $x_p$ is the free proton fraction,
$A_{ik}$ is the coefficient of $i \rightarrow k$ transition ($A_{ii}=0$),
$\alpha_i$ is the coefficient of
recombination to state $i$, $\beta_i$ is the coefficient of ionization
from state $i$. Value $J_i$ is the rate of uncompensated transitions from
state $i$ to the ground state of hydrogen atom. Value
$J_i$ is an independent parameter and to be described below.

In this work, the free electron concentration $N_e(t)$ and the free
proton fraction $x_p(t)$ are also independent parameters. These functions
are the solution of the cosmological recombination problem.
For the first time this problem has been solved by Peebles [3], 
Zel'dovich et al. [4] for hydrogen primordial plasma. 
The problem of the recombination
of the hydrogen-helium primordial plasma has been solved by Seager et al. 
[5] who used the model based on [3].

We have reproduced the results of investigations [3] and [5] (see Fig. 3). 

\Figure{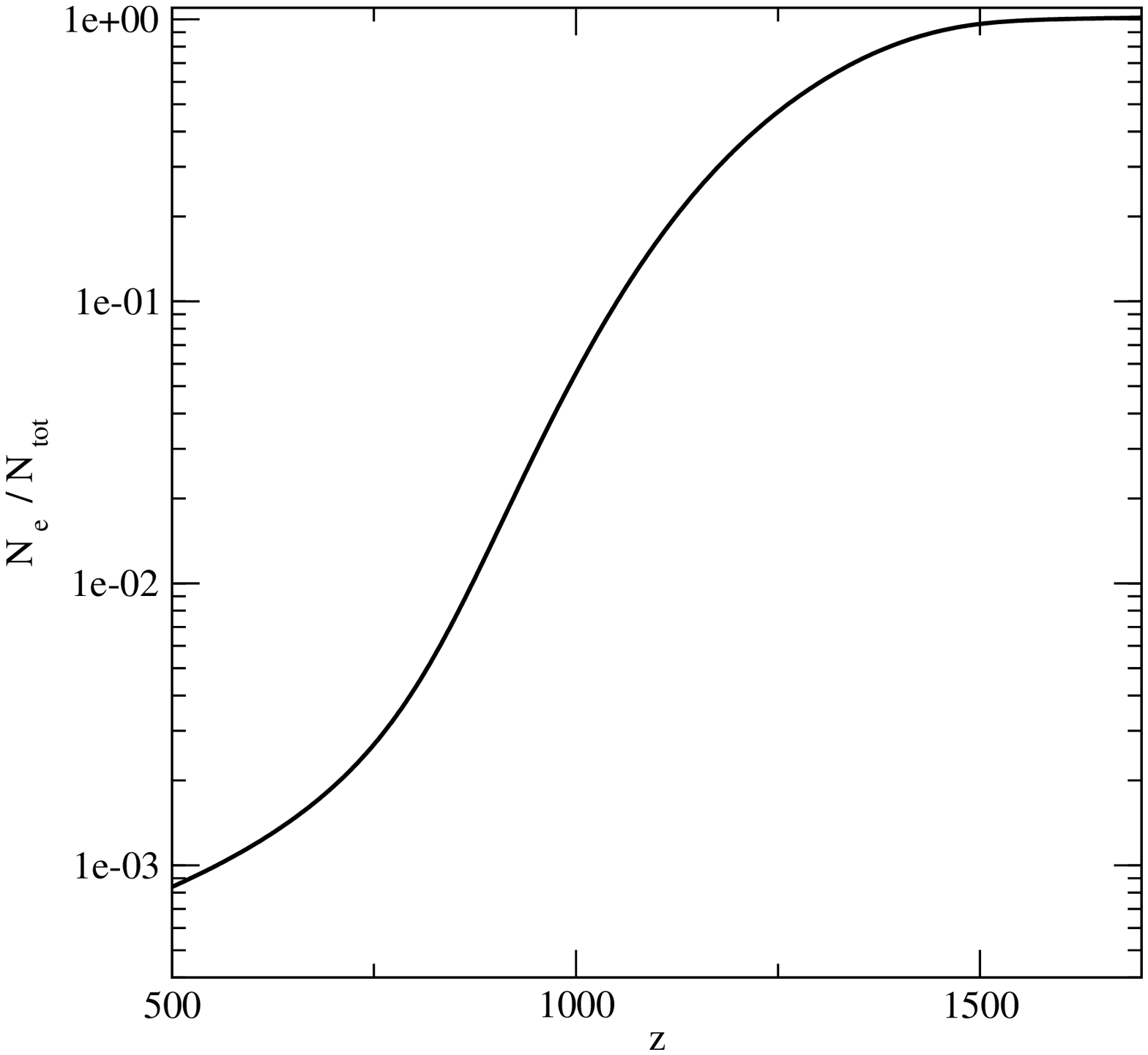} 
{The ionization fraction of the hydrogen plasma as a function of redshift}
{532}{532}

To solve equations (\ref{init_kin_eq1}) the following symbols are 
convenient: $X$ is the vector of the values $x_i$,
$B$ is the vector of the values $b_i=\alpha_i N_{e} x_p$, $J$ is the
vector of the values $J_i$, $Q$ is the transition matrix which has
the elements given by the following expressions:
\begin{equation}
Q_{ik}=A_{ki},~i\ne k;~~~~
Q_{ii}=- \left(\sum_{k=1}^{\infty}A_{ik} + \beta_i\right)
\end{equation}

These definitions allow us to rewrite equations (\ref{init_kin_eq1}):
\begin{equation}
Q(t)X-J(t)+B(t)=0
\label{matr_kin_eq}
\end{equation}
The solution of this equation can be presented in the form $X=X^0+\Delta X$,
where $X^0$ is the equilibrium values of the populations given by the
Boltzmann distribution, $\Delta X$ is the corrections concerned with the
uncompensated transitions into the ground state.
Value $X^0$ is the solution of the equation
\begin{equation}
Q(t)X^0+B(t)=0,
\label{matr_eq1}
\end{equation}
then $\Delta X$ is the solution of the equation
\begin{equation}
Q(t)\Delta X = J(t).
\label{matr_eq2}
\end{equation}
The explicit form of $\Delta X$ value is
\begin{equation}
\Delta X=Q^{-1}(t)J(t)
\label{sol_eq2}
\end{equation}
The method mentioned above (equations \ref{init_kin_eq1} -
\ref{sol_eq2}) requires preliminary determination of $J$ value.
The exact calculation of this value demands of the joint modelling of:
\\1. The recombination process.
\\2. The population behavior, taking two-quantum decay of
high-excited states into account.
\\3. The Lyman-radiation transport for each line $n \rightarrow 1$.

In this paper we used the following approximation: 
all uncompensated transitions
occur only from the first excited state due to two-quantum decay of state 2S 
and L$_\alpha$-quantum out of a line profile ([3], [4]).
Thus we can write $J_i=|\dot x_p|\delta_{1i}$ ($\delta_{ik}$ is the 
Kroneker symbol).

This definition allows us to write the following expression for the population
corrections:
\begin{equation}
\Delta x_i = (Q^{-1})_{i1}|\dot x_p|
\label{sol_eq3}
\end{equation}
System (\ref{matr_eq2}) has been solved by the Gauss method with the
choice of the leading element.
The equilibrium values have been determined by the
Boltzmann distribution.

To show a convergence of calculations with a variation of the level number, 
we used three models of the hydrogen atom: 40-level, 80-level, and 
160-level. The convergence is quite satisfactory for our aim (see Fig.7).
The standard calculations are performed for 80-level model.

\subsection{Description of radiation behavior}
In order to describe the radiation behavior, we used the kinetic
equation for radiation:
\begin{eqnarray}
{\partial \eta \over \partial t} - H\nu {\partial \eta \over \partial \nu}=
\;\;\;\;\;\;\;\;\;\;\;\;\;\;\;\;\;\;\;\;\;\;\;\;\;\;\;\;\;\;\;\;&\nonumber
\\={c^3 \over 8 \pi\nu^2}N_{tot}\sum_{i=1}^{\infty}\sum_{k=0}^{i-1}
(A_{ik} x_i-A_{ki}x_k)\Psi_{ik}(\nu)&
\label{rad_transfer_eq}
\end{eqnarray}
where $\eta(t,\nu)$ is the number of photon per mode as a function of
time and frequency, $\Psi_{ik} (\nu)$ is the line profile normalized by
the expression $\int\Psi_{ik} (\nu) d \nu =1$.

The function $\Psi_{ik}(\nu)$ has the maximum at frequency $\nu=\nu_{ik}$
and differs noticeably from zero in the frequency range $[\nu_d,\nu_u]$,~
where $\nu_d\simeq \nu_{ik}-3\Delta \nu_D$,
~$\nu_u\simeq\nu_{ik}+3\Delta \nu_D$,
~$\Delta \nu_D\simeq3\cdot 10^{-5}\nu_{ik}$ is the thermal line width.

The solution of the equation (\ref{rad_transfer_eq}) can be presented as:
\begin{equation}
\eta(t,\nu)=\eta_0(t,\nu)+\sum_{i=2}^{\infty}\sum_{k=1}^i
\Delta\eta_{ik}(t,\nu)
\end{equation}
where $\eta_0(t,\nu)$ is the solution of the homogeneous equation
corresponding to equation (\ref{rad_transfer_eq}),
$\Delta\eta_{ik}(t,\nu)$ is the partial solution of the equation:
\begin{eqnarray}
{\partial \Delta\eta_{ik} \over \partial t}
- H\nu {\partial \Delta\eta_{ik} \over \partial \nu}=&&\nonumber
\\={c^3 \over 8 \pi\nu^2}N_{tot}
(A_{ik} x_i-A_{ki}x_k)\Psi_{ik}(\nu)&&
\label{alone_line1}
\end{eqnarray}
The property of the equation (\ref{rad_transfer_eq}) allows
us to consider the equation for each line separately then to sum up 
the partial solutions.

The problem of the behavior of the non-equilibrium radiation can be
solved in two stage considered below.

\subsubsection{Formation of non-equilibrium radiation}
To research the formation of non-equilibrium radiation, we considered
the frequency range $[\nu_d,\nu_u]$ close to central frequency $\nu_{ik}$
 of the line.
The quasi-stationary approximation in this range ([3]) is:
\begin{equation}
- H\nu {\partial \Delta\eta_{ik}^s \over \partial \nu}
={c^3 \over 8 \pi\nu^2}N_{tot}
(A_{ik} x_i-A_{ki}x_k)\Psi_{ik}(\nu)
\label{alone_line2}
\end{equation}
where $\Delta\eta_{ik}^s(t,\nu)$ is the partial solution in the range
considered. This solution depends on time parametrically.
The value $\Delta\eta_{ik}^s(t,\nu)$ at the lower limit of the range 
considered sets
the boundary condition for the further evolution of the non-equilibrium
radiation: $\Delta\eta_{ik}^d(t)=\Delta\eta_{ik}^s(\nu_d,t)$.
This value can be obtained by the formula based on the general solution of
the radiation transport equation in the case of the interaction between 
radiation and matter at different temperatures ([6]):
\begin{equation}
\Delta\eta_{ik}^d= (\eta_m - \eta_0 (\nu_{ik}))(1 - \exp(-\tau_{ik}))
\label{eta_source}
\end{equation}
where value $\eta_m=1/(x_k i^2/x_i k^2-1)$ describes the radiation of
matter, $\eta _0$ is the photon number per mode in the flow of the incident
quanta. In our case $\eta_0$ is given by the Planck distribution.
The value $\tau_{ik}$ is the optical depth in the line $i\rightarrow k$ 
([7], [8]):
\begin{equation}
\tau_{ik}={c^3 \over 8\pi\nu_{ik}^3}N_{tot}{A^s_{ik} \over H}({i^2
\over k^2}x_k - x_i)
\label{optical_depth}
\end{equation}
where $A^s_{ik}$ is the Einstein coefficient of the spontaneous
transition ([9], [10]). The formula (\ref{optical_depth}) is 
obtained by means of the absorbtion coefficient formula of 
line $i \rightarrow k$ ([11]).

We can linearize expression (\ref{eta_source}) over the optical depth, if it
is much less than one ($\tau_{ik}\ll 1$):
\begin{equation}
\Delta\eta_{ik}^d=(\eta_m - \eta_0 (\nu_{ik}))\tau_{ik}
\end{equation}
and using (\ref{optical_depth}) it can be written:
\begin{equation}
\Delta\eta_{ik}^d
={c^3 \over 8 \pi \nu_{ik}^3} N_{tot}(z) H^{-1}(z)(A_{ik}x_i-A_{ki}x_k)
\label{BBD_appr1}
\end{equation}
The following definition is convenient:
\begin{equation}
\theta_{ik}=J_1^{-1}(A_{ik}x_i-A_{ki}x_k)
\label{ert_matr_defin1}
\end{equation}
It can be rewritten (in our model):
\begin{equation}
\theta_{ik}=|\dot x_p|^{-1}(A_{ik}\Delta x_i-A_{ki}\Delta x_k)
\label{ert_matr_defin2}
\end{equation}
Taking into account the relation $dz/dt=-H(z)(1+z)$ and expression
(\ref{ert_matr_defin2}), formula (\ref{BBD_appr1})
can be rewritten in the form:
\begin{equation}
\Delta\eta_{ik}^d =
{c^3 \over 8 \pi \nu_{ik}^3} N_{tot}(z) (1+z){dx_p \over dz}\theta_{ik}
\label{BBD_appr2}
\end{equation}
This formula allows us to obtain the following (in the case
$h\nu_{ik}\ll k_BT$):
\begin{equation}
{\Delta T \over T}={c^3 \over 8 \pi \nu_{ik}^3}
(1 + z) {dx_p \over dz}{h\nu_{ik} \over k_B T(z)}N_{tot}(z) \theta_{ik}
\label{delta_T}
\end{equation}
For the first time this formula has been obtained in the 
paper [1] in somewhat other form.

We can estimate $dx_p / dz$ on its averaged value
$\overline{dx_p/dz}=1/\Delta z$. Then we obtain formula (\ref{estimate2}).

The element of the ERT-matrix $\theta_{ik}$ means 
the averaged number of photons emitted in line $i\rightarrow k$ per one act of 
uncompensated recombination.

Taking definition (\ref{ert_matr_defin2}) and expression (\ref{sol_eq3})
into account, the elements of the ERT-matrix are given by the formula:
\begin{equation}
\theta_{ik}=(A_{ik}(Q^{-1})_{i1}-A_{ki}(Q^{-1})_{k1})
\label{ert_matr2}
\end{equation}
It shows that the elements of the ERT-matrix can be calculated
independently with the solution of the cosmological recombination problem. 

In paper [1] the calculation of the ERT-matrix 
was necessary because of the approach accepted. In paper [2]
the ERT-values were the main result. To compare our result with that of 
paper [2], ERT-matrix has been calculated in this work (Fig. 4, Fig. 5). 
The Burgin's results [2] have been reproduced 
with relative accuracy within 1\%.

\Figure{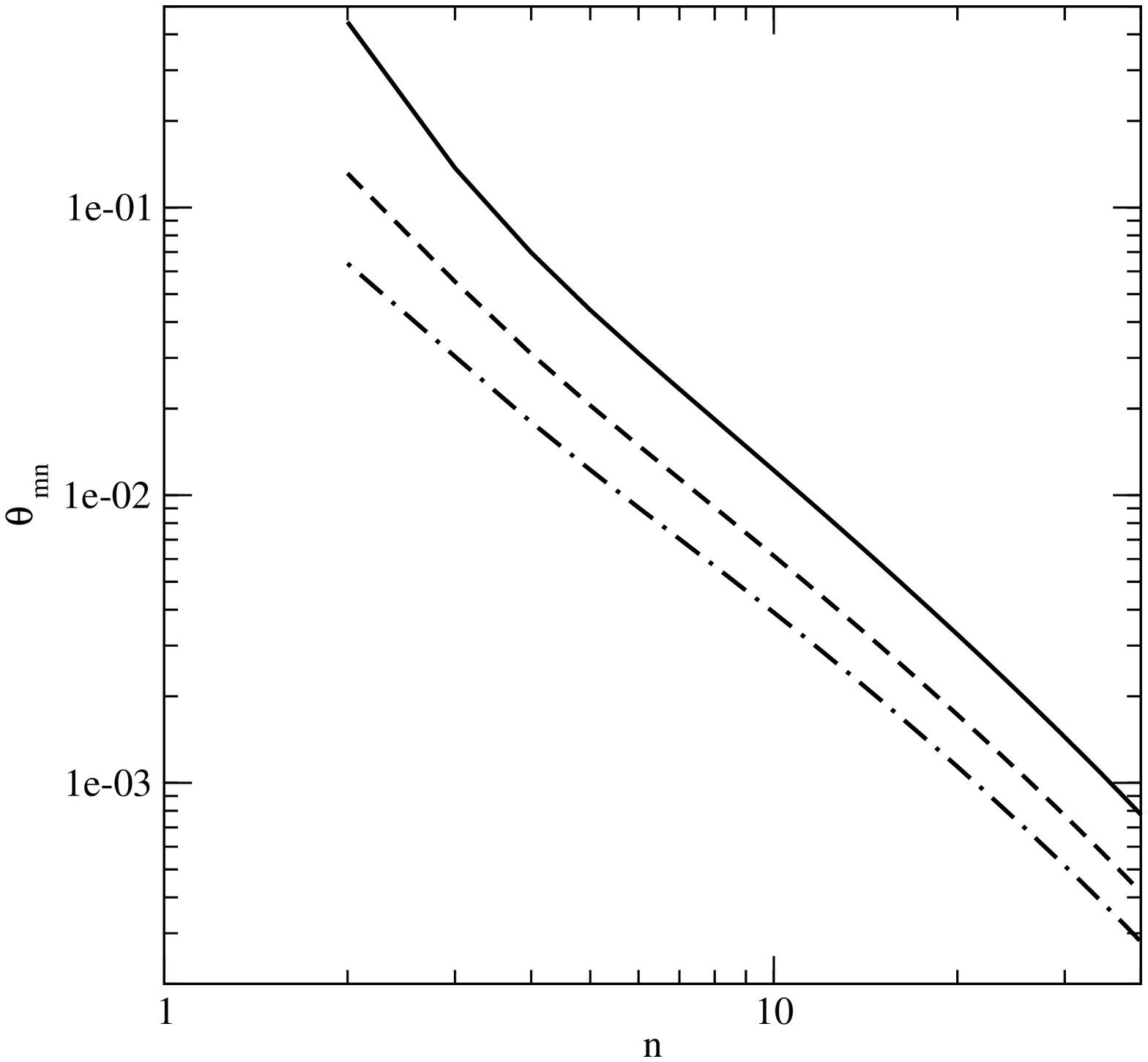} 
       {The ERT-elements as functions of the principal quantum number $n$ of 
       lower level: the solid line corresponds to transition with $m=n+1$, 
       dashed -- $m=n+2$, dashed-dot -- $m=n+3$}{540}{532}
\Figure{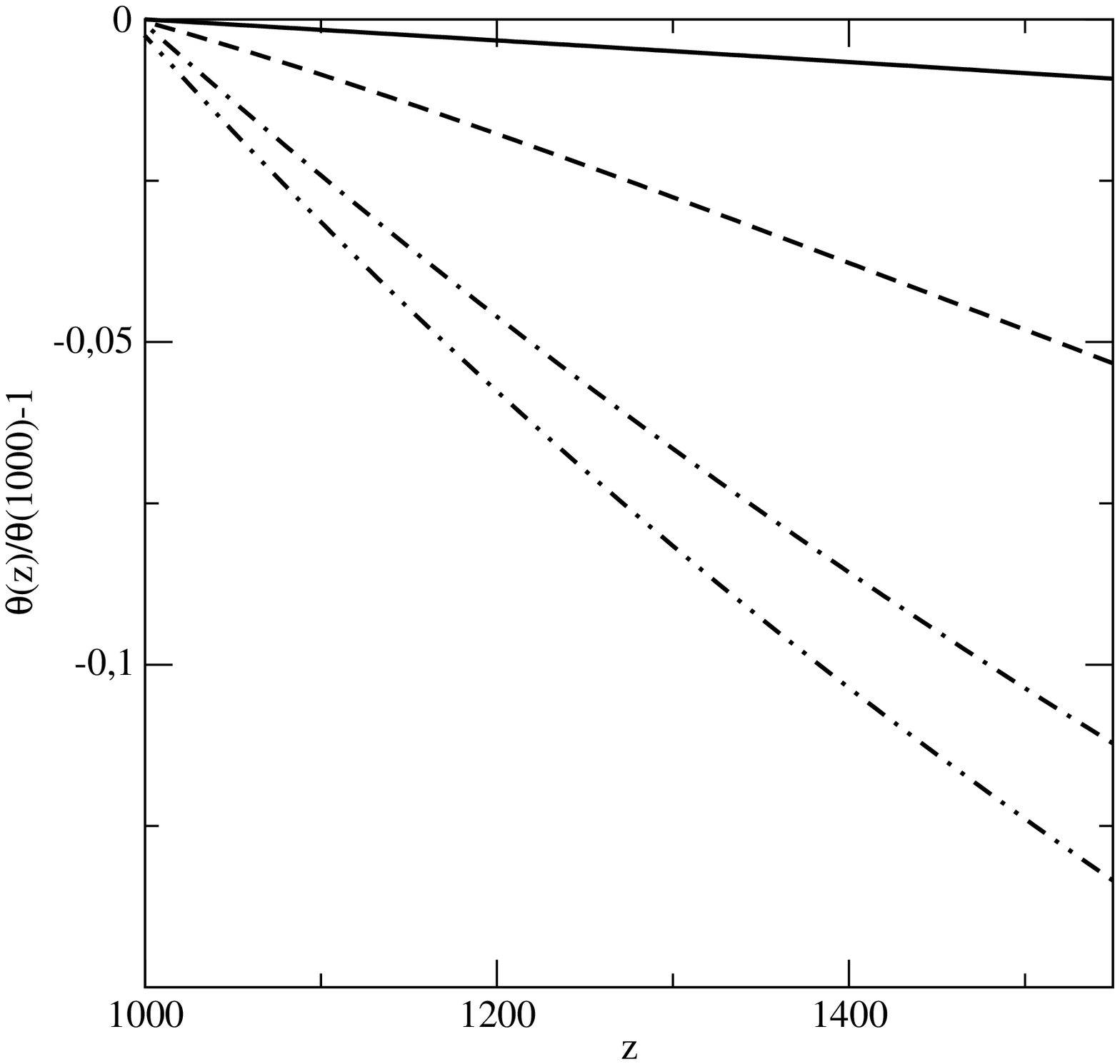} 
       {The relative change of ERT-elements during the recombination epoch: 
       from top to bottom 
       the transitions are showed: $3\rightarrow 2$,  $11\rightarrow 10$, 
        $21\rightarrow 20$,  $31\rightarrow 30$}{540}{532}

\subsubsection{Radiation transport in homogeneous expanding
environment}
To investigate the radiation transport in the homogeneous expanding
environment, we considered the frequency range $[0,\nu_d]$.

Due to the properties of function $\Psi_{ik}(\nu)$,
 equation (\ref{rad_transfer_eq}) in this range is:
\begin{equation}
{\partial \over \partial t} \Delta\eta_{ik}
- H\nu{\partial \over \partial \nu}\Delta\eta_{ik} = 0
\label{alone_line3}
\end{equation}
where $\Delta\eta_{ik}(t,\nu)$ is the partial solution
with the boundary condition $\Delta\eta_{ik}(t,\nu_{ik})=\Delta\eta_{ik}^d(t)$
(taking $\nu_{ik}\simeq \nu_d$ into account).

Changing from time-functions to redshift-functions,
 equation (\ref{alone_line3}) takes the form:
\begin{equation}
(1+z){\partial \over \partial z} \Delta\eta_{ik} +
\nu{\partial \over \partial \nu}\Delta\eta_{ik} =0
\label{transport2}
\end{equation}
with the boundary condition
$\Delta\eta_{ik}(z,\nu_{ik})=\Delta\eta_{ik}^d(z)$.

The solution of the equation (\ref{transport2}) is:
\begin{equation}
\Delta\eta_{ik}(z,\nu)=\Delta\eta_{ik}^d((1+z)\nu_{ik}/\nu-1)
\end{equation}

The number of photons per mode at frequency $\nu$ in the
epoch with redshift $z$ is given by the formula:
\begin{equation}
\eta(z,\nu)=\eta_0(z,\nu)+\sum_{i=2}^{\infty}\sum_{k=1}^i
\Delta\eta_{ik}(z,\nu)
\end{equation}
The relative distortion $\Delta T / T = (T_{ex}-T)/T$
of the CMBR temperature is calculated by the formula:
\begin{equation}
T_{ex}(z,\nu)={ h \nu \over k_B \ln(1 / \eta (z,\nu) + 1)}
\end{equation}

\section{Features of observation}
Many lines cannot be registered because the CMBR
 is contaminated by quasar radiation and IR-source radiation ([12]). 
For example, the strongest expected
line L$_\alpha$ is invisible because of dust radiation. 
Taking into account this background, it is reasonable to calculate
the recombination radiation concerned with the transitions with
the principal quantum number over ten. The spectrum of this
recombination radiation is located in the frequency range
observed with the detectors for CMBR measurements.

\section{Results}
The main result of this paper is the relative CMBR temperature distortion 
as a function of frequency. This result is shown in Fig.6. 
We calculated the CMBR distortion with the various sets 
of the cosmological parameters $\Omega_m$ and $\Omega_b$: 
$\{\Omega_m=0.3;~h^2\Omega_b=0.014,\;0.018,\;0.022\}$ and 
$\{\Omega_m=0.1,\;0.3,\;0.5;~h^2\Omega_b=\;0.018\}$
\Figure{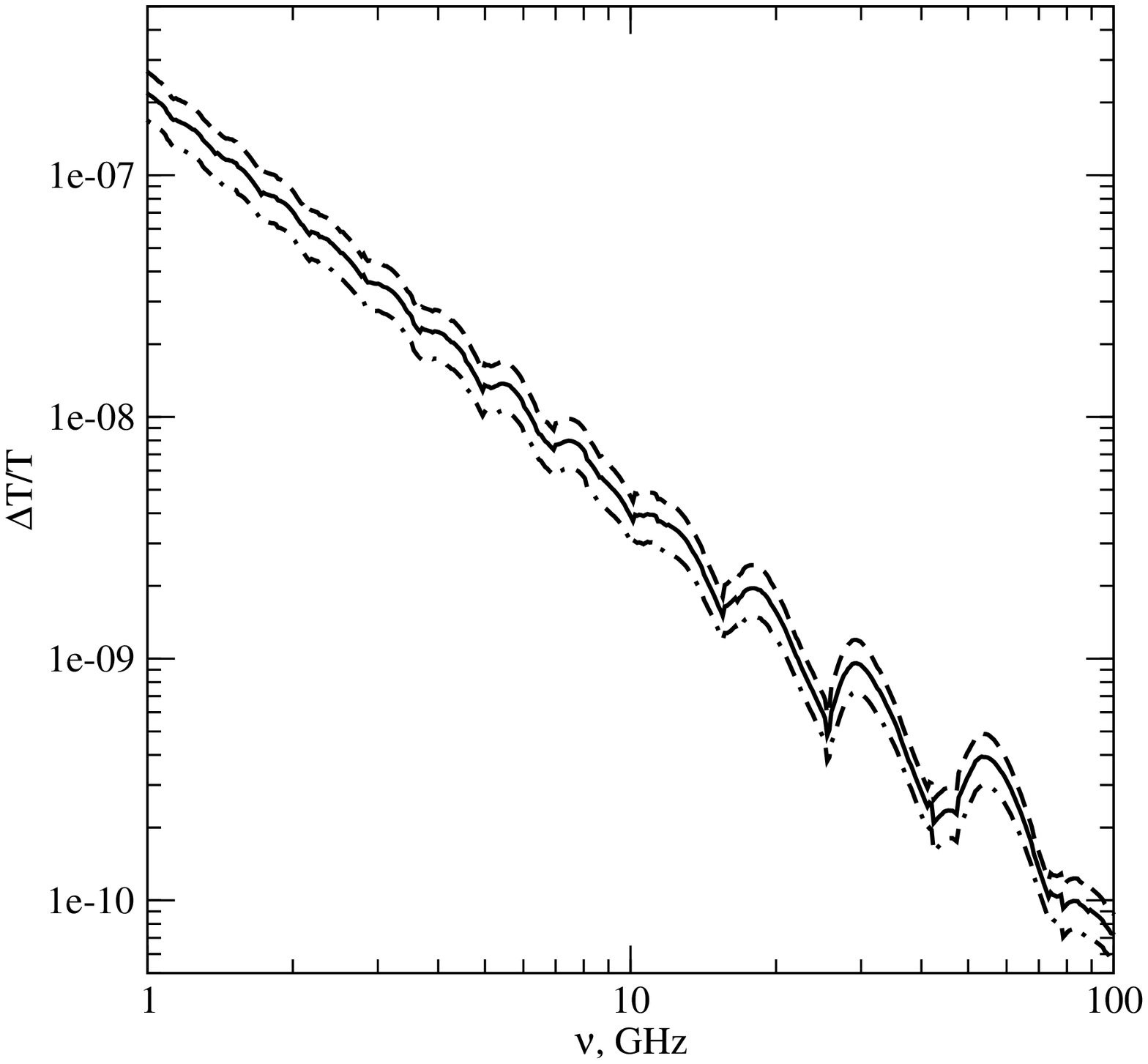}  
 {The relative CMBR temperature distortion as a function of frequency. The 
curves correspond to $h^2\Omega_b=0.014;0.018;0.022$ from bottom 
to top. All curves were calculated at $\Omega_m=0.3$}{540}{532}

The result of $\Omega_m$-variation is not presented graphically because of 
weak dependence of CMBR distortion on $\Omega_m$-value. This fact can 
be explained for the optically thin lines (in our case $\tau<10^{-4}$) by 
formula (\ref{BBD_appr2}). It displays 
that, the CMBR distortion $\Delta\eta$ depends on $\Omega_m$ by only 
the ionization fraction $x_p(z)$ which vary with $\Omega_m$ weakly.

The $\Omega_b$-variation shows that CMBR temperature distortion is 
directly proportional to the baryonic matter density $\Omega_b$ (Fig.6).

The maximum value of the relative CMBR distortion in the observed range 
is less than $3\cdot 10^{-7}$. This value is to be observed at the low 
frequencies. 

To resolve the distortion corresponding to different values $\Omega_b$, 
it is necessary to measure the CMBR temperature with absolute accuracy 
$\sim$ 1 nK. 

\Figure{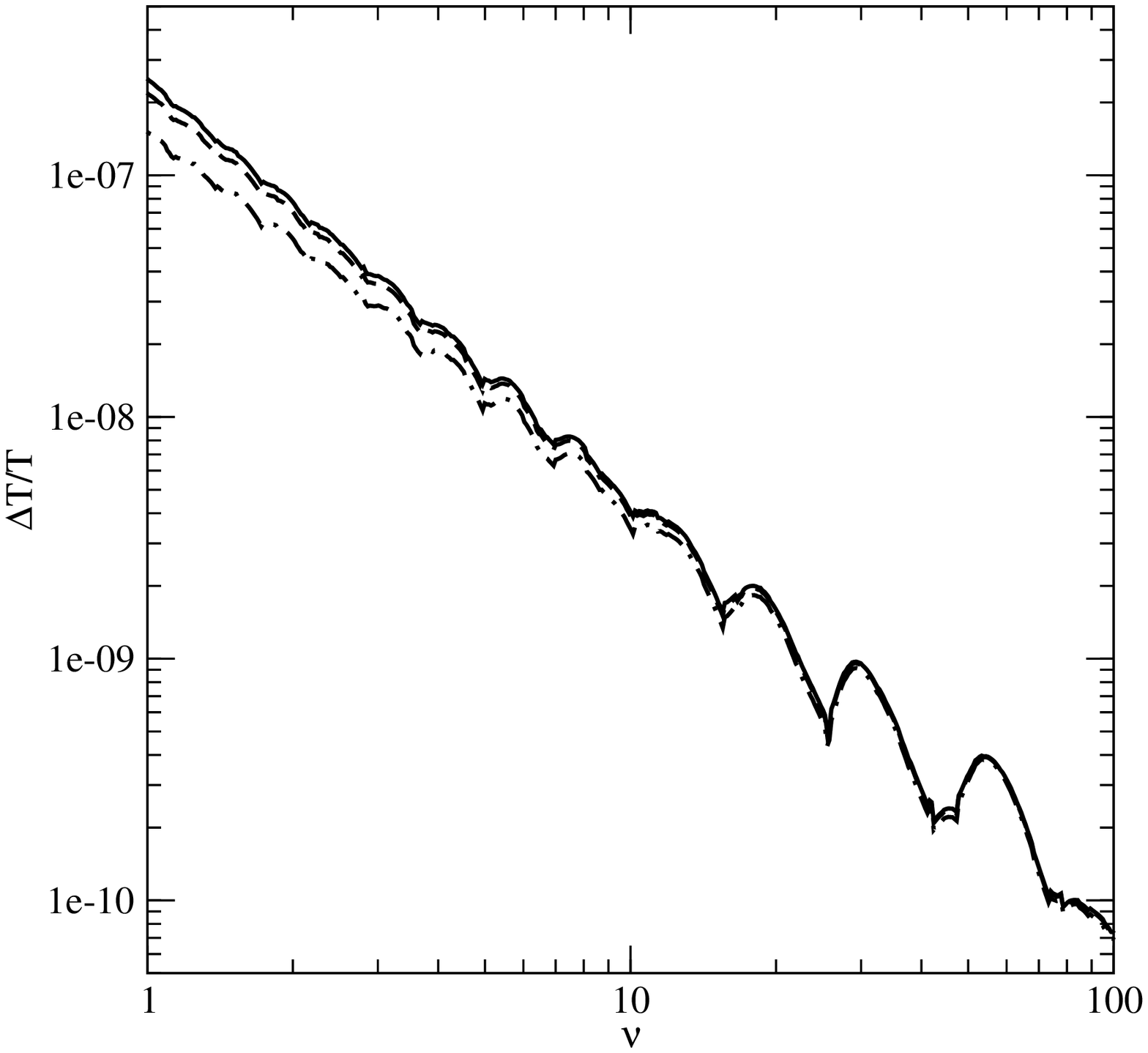}  
 {The distortion at variation of the level number of hydrogen atom. The 
curves correspond to $n_{max}=40,\;80,\;160$ from bottom 
to top. All curves were calculated at $\Omega_m=0.3, h^2\Omega_b=0.018$}
{540}{532}

\Acknow
{This work is performed with support of the RFBR grants (02-02-16278-a, 
03-07-90200-v) and the student grant (2003) of the Dynasty Foundation.}


\small
\begin{thebibliography}{99}
\bibitem{1}
   I. N. Bernstein et al., {\it Sov. Astron. J.\/} {\bf 54}, 727 (1977).
\bibitem{2}
   M. S. Burgin, {\it Astron. zhurnal\/}, {\bf 80}, 771 (2003) 
\bibitem{3}
   P. J. Peebles, {\it ApJ.\/} {\bf 153}, 1 (1968).
\bibitem{4}
   Ya. B. Zeldovich et al., {\it Sov. Astron. J.\/} {\bf 55}, 287 (1968).
\bibitem{5}
    S. Seager et al., ``A new calculation of the recombination epoch'', 
    astro-ph/9909275; 
\bibitem{6}
     L. Spitzer, ``Physical processes in the interstellar medium'',
     publishers ``Mir'', Moscow, 1981.
\bibitem{7}
   S. Seager et al., {\it ApJ. Suppl.\/} {\bf 128}, 407 (2000).
\bibitem{8}
    E. Kholupenko et al., PhTI preprint 1758 (2002)
\bibitem{9}
     V. B. Berestetzkii et al., ``Quantum electrodynamics'',
     publishers ``Nauka'', Moscow, 1980.
\bibitem{10}
     L. C. Johnson, {\it AJ.\/} {\bf 174}, 227 (1972).
\bibitem{11}
     K. Leng, ``Astrophysical formulae'',
     publishers ``Mir'', Moscow, 1978.
\bibitem{12}
     M. S. Longeir, R. A. Syunyaev, {\it UFN\/} {\bf 153}, 
     41 (1971).
\end{thebibliography}
\end{document}